# Halogen Doped Electronic Properties of 2D ZnO: A First Principles Study

Hossain Mansur Resalat Faruque[1], Kamal Hosen[2], A. S. M. Jannatul Islam[3], and Md. Sherajul Islam[4]
Department of Electrical and Electronic Engineering, Khulna University of Engineering & Technology
Khulna-9203, Bangladesh
resalat_f@yahoo.com[1], raiyankamal1721.kuet@gmail.com[2], jannatul@eee.kuet.ac.bd[3],
sheraj_kuet@eee.kuet.ac.bd[4]

*Abstract*—In recent times, two dimensional (2D) ZnO has attracted a great attention in the field of nano-research due to its extraordinary electronic, thermal and optical properties. In this paper, we have explored the effects of halogen impurity doping such as F, Cl, and Br atoms on the electronic properties of 2D ZnO using first principles calculation. The pristine 2D ZnO exhibits a semiconducting behavior with a direct bandgap of 1.67 eV on the Γ point. However, when impurities such as F, Cl, or Br atoms are introduced, the 2D ZnO shows semi-metallic behavior with almost zero bandgap. It is perceived that, owing to the introduction of F impurity, the zero bandgap is exhibited at the K point of the electronic band structure. However, in the case of Cl and Br impurities, the nearly zero bandgap is observed elsewhere rather than on the K point. Moreover, due to the introduction of impurity atoms, the Fermi level also shifted towards the conduction band (CB) suggesting an increase of the carrier concentration in the density of states (DOS) results. These findings might be very much beneficial when doping effect, especially halogen impurity doping is considered to modulate the electronic properties of 2D ZnO in the near future.

## I. INTRODUCTION

Recent technological advancement triggered by the successful exfoliation of graphene, a two dimensional (2D) allotrope of carbon atoms [1], has created a profuse research interest in the field of nanotechnology. The amazing electrical, mechanical, and thermal properties [2, 3] of graphene inspired the scientists and researchers to discover the characteristics of other 2D materials [4] specially binary compound systems. Therefore, in recent times, group II-VI and III-IV based 2D compounds, achieving a great popularity in nanoresearch owing to their tunable behaviors [5, 6] and applicability in various realistic uses. Unlike graphene, binary compounds have well-defined thermal, chemical and structural stsbility as well as direct electronic bandgap, which is very much essential to fulfil the requirements of modern device fabrication.

Of late, 2D ZnO, a group II-VI binary compound of Zn and O atoms has attracted a prolific research attention due to its intriguing electronic bandgap (~1.68 eV), large excitonic binding energy (~60 meV) and anomalously decreasing thermal conductivity [7] properties. Several studies predicted that 2D ZnO has a honeycomb crystal structure analogous to graphene [8] and can shows superior stability compared to other 2D materials. Besides, nanosheets [9], nanowires [10], and nanotubes [11] of ZnO have already been synthesized and it has been found that this nanomaterial has a high potentiality to be applied in nanoelectronic and optoelectronic applications including solar cells, light emitting diodes (LED), LASERs, etc. [12]. However, to control the device efficiently in different circumstances, multifunctional applications require appropriate tunable electronic properties. In the practical world, it is almost impossible to attain a pure material which is free from impurity and it is a general practice to introduce the impurity in the material for changing its electronic properties smoothly. The electronic properties of nanomaterials also considerably depend on the types of applied impurities [13]. Semiconducting, metallic, semimetallic, insulating etc. behaviors can be easily achieved with the help of appropriate impurities.

Moreover, introduction of Halogen impurities in 2D materials creates an unimaginable research corner presently. The electronic, sensing, electrochemical, and mechanical properties of 2D systems such as graphene, TMDs [16] are greatly affected by the Halogen doping. Halogen impurity with Mg and P atoms can also be used as the performance tuner of the optoelectronic devices like solar cells, lasers, photodetectors etc. [15] Consequently, as 2D ZnO is structurally analogous to graphene and other 2D systems, we expect intriguing results from this material with halogen doping. However, to the best of our knowledge, there is no study on the electronic properties of 2D ZnO considering the effect of Halogen impurities. To ensure the application needs and efficient operation of ZnO based nanoelectronics, impact of Halogen impurities on the electronic properties, thus should be properly understood. In this paper, we have explored the effect of group VII impurities such as Cl, F, and Br atoms on the electronic properties of 2D ZnO using the first principle calculation based on density functional theory (DFT). The electronic band structure, density of states (DOS), and projected density of states (PDOS) have been calculated systematically. Extraordinary results are found from this study which will be beneficial to design ZnO based nanodevices in the near future.

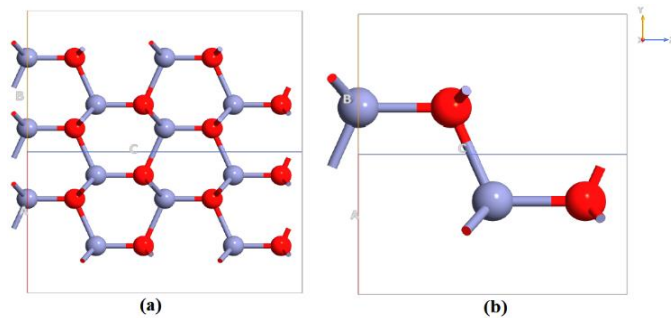

Figure 1. (a) Electronic structure of 2D hexagonal ZnO sheet, (b) unit cell of 2D hexagonal ZnO, where red stands for O and gray stands for Zn atom.

## II. COMPUTATIONAL METHOD

In this work, the first principle calculations were conducted within the framework of density functional theory (DFT) [3]. The quantum espresso package was used to conduct the electronic property calculations of 2D ZnO. The linear difference approximation (LDA) was used as the exchange-correlation function to solve the Kohn Sham equation [5]. For representing the atomic interaction ultra-soft pseudo-potential was used. The electronic structure of the hexagonal 2D ZnO sheet is shown in Fig. 1 (a). Each unit cell consists of two Zn atom and two O atom and is represented in Fig. 1 (b). The Zn-O bond length is considered to be d=1.8 Å. Besides, the Zn-O-Zn bong angle is considered to be 120º. In the unit cell, the O atom is replaced by the F, Cl, and Br atom, which represented in Fig 2. $Zn-4s^2$, $O-2s^22p^4$, $F-2s^22p^5$, $Cl-3s^23p^5$ and $Br-4s^24p^5$ have been considered as valence electron for the structures. As 2D ZnO has a hexagonal structure, hexagonal Brillouin zone (BZ) with Γ-K-M-Γ directions is considered in the calculation of the band structure. A 12×12×1 Monkhors-Pack grid was considered during the BZ sampling for the geometric optimization. For DOS and PDOS calculations, a denser K point sampling of 36×36×1 grid was considered to ensure more accuracy. During each consecutive self-consistency step, the change in energy and the estimated energy error was considered to be less than $10^{-4}$ Rydeberg and $10^{-6}$ Rydeberg, respectively. Plane wave basis with a kinetic energy cut off value of 30 eV for wave function and 250 eV for charge density and potential were used to ensure convergence. The Fermi level was set at zero for simplification.

## III. RESULTS AND DISCUSSION

To understand the effects of the group VII atoms, at first the electronic properties of pure 2D ZnO is analyzed. Fig. 3 represents the band structure and corresponding DOS of pure 2D ZnO sheet. From the band structure, it is evident that the pure 2D ZnO has a direct band gap of 1.67 eV at the Γ point. From the DOS calculation, it can be concluded that the valence band (VB) contribution is much higher than the conduction band (CB) contribution. Therefore, it exhibits p-type semiconducting property. It also has moderate DOS peaks, which indicates moderate carrier concentration. The corresponding PDOS calculation is represented in Fig. 4. The d orbital of Zn and p orbital O has a high effect on the VB, whereas, the p orbital of both Zn and O has a high effect on the CB. The s orbital of Zn and O has negligible effect on both the CB and VB.

The electronic properties can be changed when an impurity atom is introduced. The band structure & the corresponding DOS calculation of 2D ZnO with F impurity is shown in Fig. 5. It is evident that the VB crosses the Fermi level, but does not overlap with CB and at K point the band gap is nearly zero, which indicates semi metallic behavior. This is due to fact that the interaction of F atoms with the Zn atom creates a magnetic moment, which weakly breaks the time reversal symmetry (TRS) of the system and lifts the Kramer degeneracy at Γ point. Therefore, at Γ point, the CB is much closer to the Fermi level compared to the case for pure 2D ZnO, which indicates a shift of the Fermi level towards the CB. The F atom has one more valence electron than the O. Hence, after forming the covalent bond with Zn it will have a more unpaired electron than O. Therefore, the carrier concentration also increases, which is evident from the large peak of the DOS calculation. From the PDOS calculation represented in Fig. 6, it is evident that p orbital of O, and d orbital of Zn has higher contribution on the VB. The p orbital of F has a moderate effect on the VB and negligible effect on the CB. On the other hand, the s and p orbital of Zn has higher contribution on the CB.

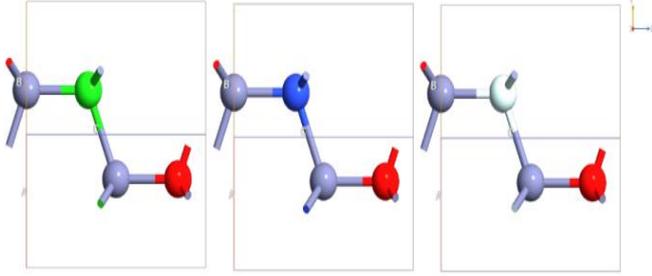

Figure 2. Electronic structure of 2D ZnO with (a) F impurity, (b) Cl impurity, and (c) Br impurity. Here green, red and white balls stands for F, Cl, and Br, respectively. The O atom in the middle is replaced by F, Cl and Br.

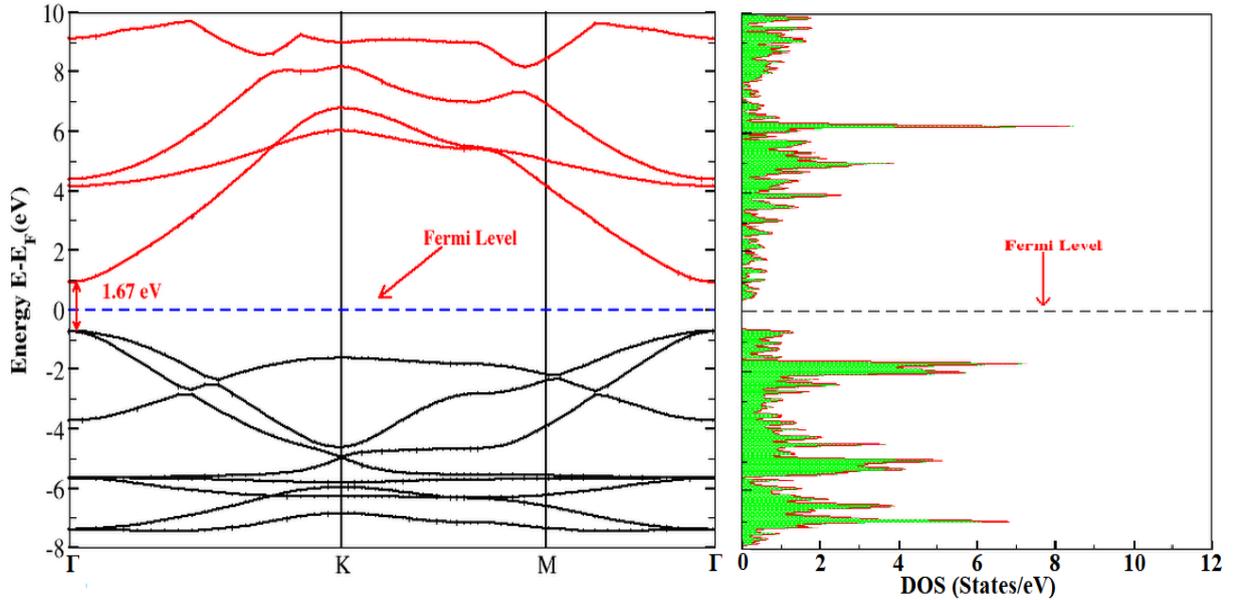

Figure 3. The band structure and corresponding DOS calculation of pure 2D ZnO.

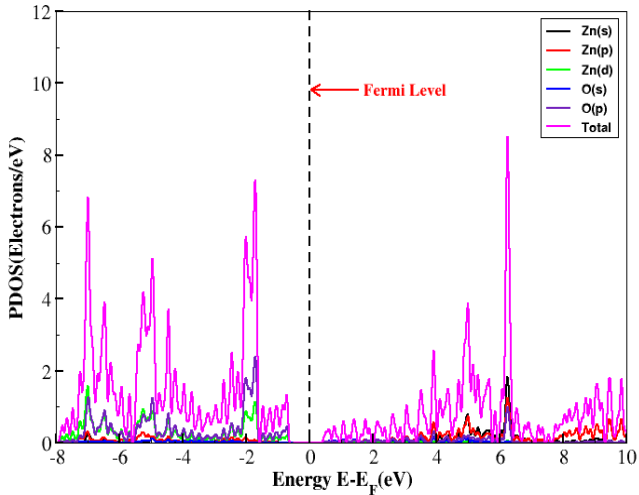

Figure 4. PDOS calculation of pure 2D ZnO.

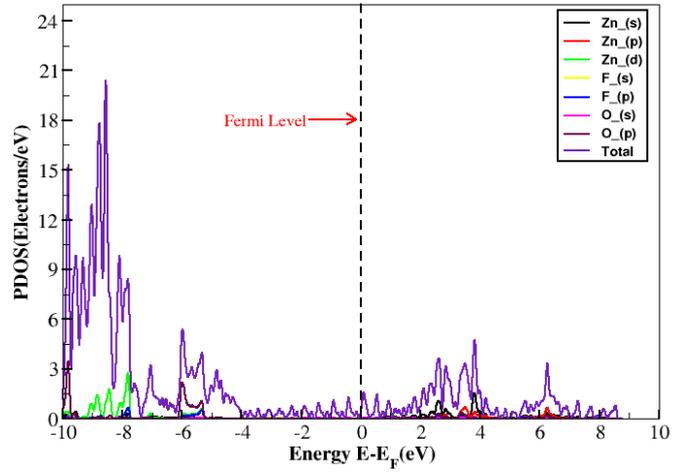

Figure 6. PDOS calculation of 2D ZnO with F impurity.

The band structure & the corresponding DOS and PDOS calculation of 2D ZnO with Cl impurity is represented in Fig. 7 and Fig. 8, respectively. From Fig. 8 it is clearly evident that the VB crosses the Fermi level, however, it does not overlap with the CB. Therefore, 2D ZnO with Cl impurity exhibits semi metallic behavior with almost zero bandgap, which is not exhibited at the K point. Similar to F atom, Cl atom also has 7 valence electrons, whereas, O has 6. Therefore, the number of unpaired electrons of Cl atoms after forming covalent bonds with Zn, will be more than O. Hence, the carrier concentration will also be higher, which is evident from the high peak and widespread DOS. Another important discovery is that for Cl impurity, the CB at Γ point is much closer to the Fermi level than in the case with F impurity. This is because the Cl has a higher electron concentration than the F atom. The interaction of Cl atoms with Zn creates a magnetic moment, which is much more than in the case with F impurity, and thus; the TRS is broken, lifting the Kramer degeneracy at Γ point, which results in shifting the Fermi level more towards the CB at the aforementioned point. From the PDOS calculation of ZnO with Cl impurity it is evident that the d orbital of Zn, p orbital of Cl, and p orbital of O has the high contribution in the VB, whereas, for CB, s orbital of Zn and p orbital of O has high contribution.

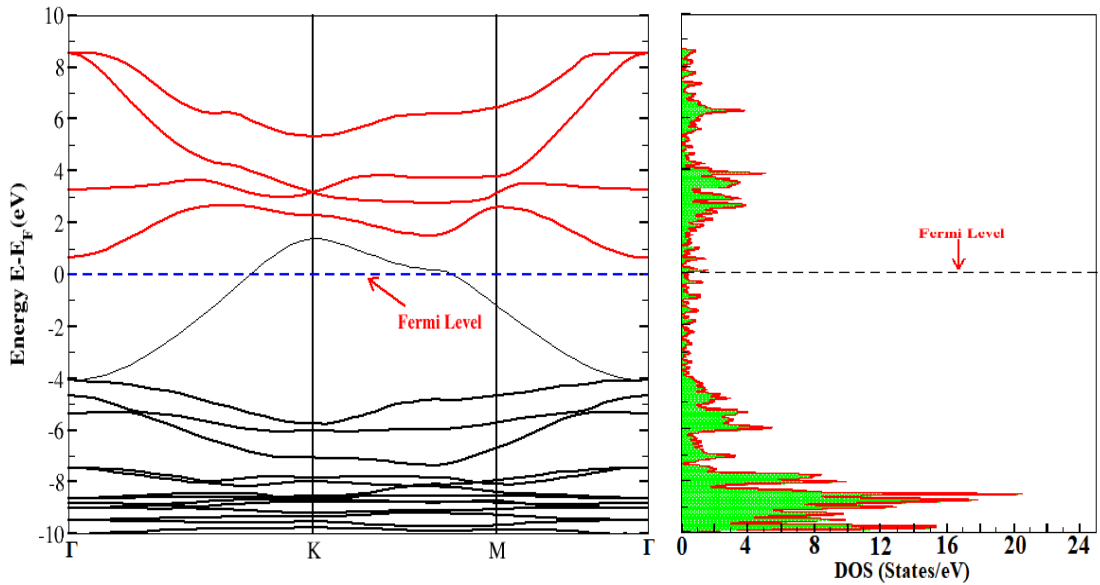

Figure 5. The band structure and corresponding DOS calculation of 2D ZnO with F impurity.

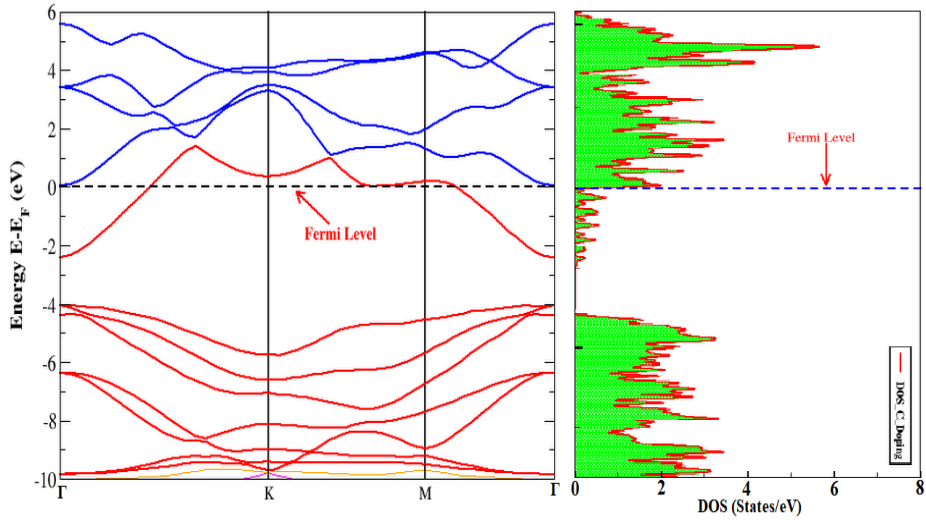

Figure 7. The band structure and corresponding DOS calculation of 2D ZnO with Cl impurity.

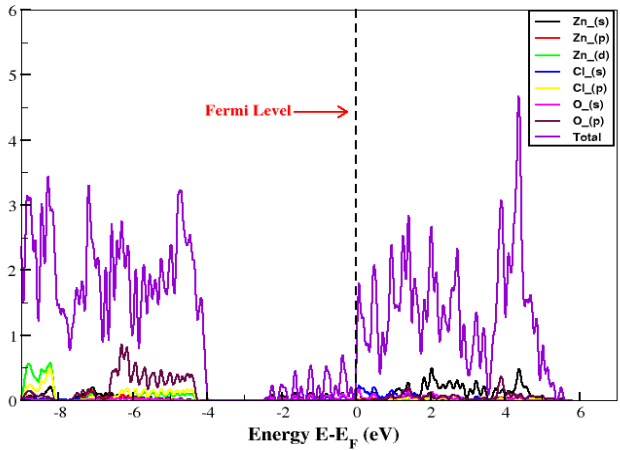

Figure 8. PDOS calculation of 2D ZnO with Cl impurity.

Fig. 9 represents the band structure & corresponding DOS of 2D ZnO with Br impurity atoms. Just like the previously discussed cases, ZnO with and Br impurity exhibits semi metallic behavior as the VB crosses the Fermi level but does not overlap it. However, the almost zero bandgap is not exhibited at K point. Br having more valence electron than O is the reason behind this result. The DOS calculation shows high peaks and widespread DOS, which means high carrier concentration. In this case, the carrier concentration is higher than the pure ZnO and ZnO with F impurity. However, it is less than ZnO with Cl atoms. Br atoms having electrons in the d orbitals may be the reason behind this result. Consequently, the Fermi level shift towards the CB is also evident in this case, even though it is less than in the case with Cl impurity. PDOS calculation in this case is represented in Fig. 10. The conducted PDOS calculation of ZnO with Br impurity suggests that the d orbital of Zn, p orbital of Br, and p orbital of O has a higher effect on the VB. On the other hand, s orbital of Zn and p orbital of O has higher contribution on the CB. The Fermi level shift towards the CB at Γ point for the introduction of different halogen impurity atoms is tabulated in Table I.

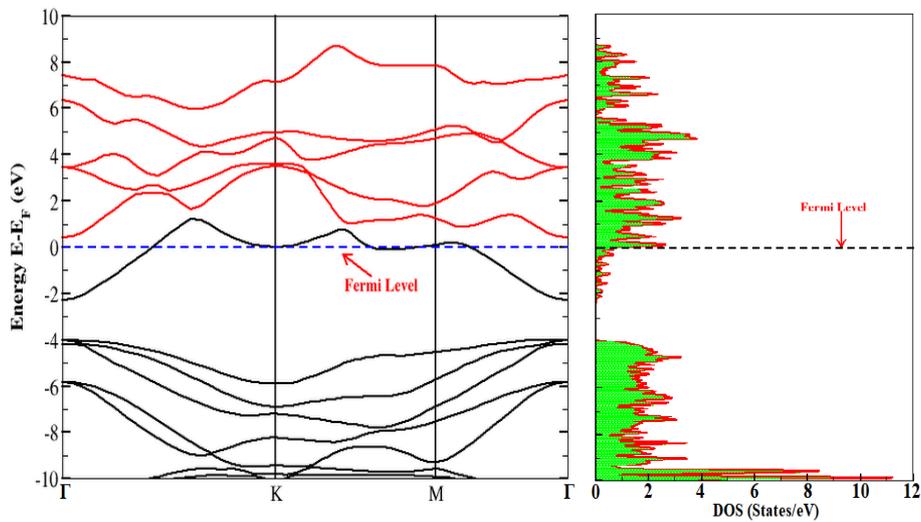

Figure 9. The band structure and corresponding DOS calculation of 2D ZnO with Br impurity.

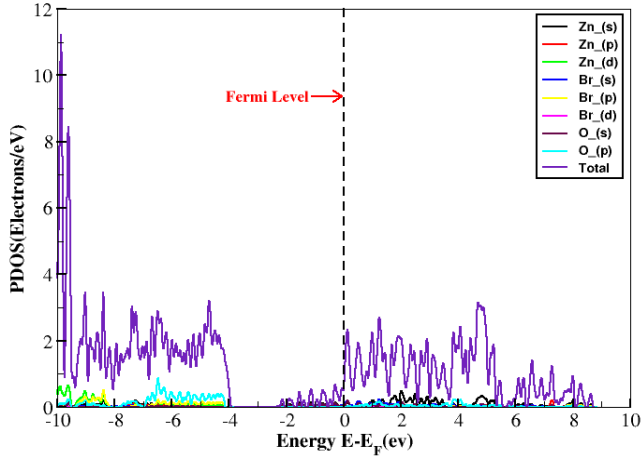

Figure 10. PDOS calculation of 2D ZnO with Br impurity.

TABLE I. SHIFT IN FERMI LEVEL TOWARDS CB FOR DIFFERENT CASES

| Case | Difference between the CB and Fermi energy at Γ point |
|---|---|
| Pure 2D ZnO | 0.89 eV |
| ZnO with F impurity | 0.56 eV |
| ZnO with Cl impurity | 0.069 eV |
| ZnO with Br impurity | 0.37 eV |

## IV. CONCLUSION

The main incentive of this work was to determine how the electronic properties of 2D ZnO change due to the introduction of impurity atoms and the reasons behind these changes. Group VII elements such as F, Cl, and Br have been used as the impurities. The pure 2D ZnO exhibited p-type semiconducting property with a direct band gap of 1.67 eV at the Γ point. For the introduction of F, Cl, and Br impurities, the ZnO exhibited semi-metallic behavior for all the cases. From the PDOS calculation, it has been observed that the impurities have a significant contribution on the VB. In all the cases the d orbital of Zn and p orbital of O has the highest contribution on VB. On the other hand, the s orbital of Zn and p orbital of O has higher contribution on CB. It's also evident that with the introduction of Halogen impurities the Fermi level shifts towards the CB due to having a number of unpaired electrons after forming covalent bonds with Zn atom. Furthermore, the carrier concentrations of 2D ZnO with the aforementioned impurities are also higher than the pure ZnO. The 2D ZnO structure with Cl impurity has the highest carrier concentration among all the structures. Therefore, the Fermi level shift towards the CB is more prevalent for the case with Cl impurity than for the other cases. From the obtained results, it can be concluded that the electronic properties of 2D ZnO can be modulated with the introduction of other impurity atoms; therefore, it has high potentiality for developing nano-devices.